\documentstyle[12pt]{article}
\begin{document}

\def\fq{FQHE~}
\def\nq{\frac{N}{q}}
\def\ne{{N_e}}

\begin{flushright}
EFI 92-47 \\
{\it Preliminary Version} \\
\end{flushright}

\bigskip
\bigskip
\begin{center}
{\bf MANY PARTICLE HAMILTONIAN FOR THE \\
FRACTIONAL QUANTUM HALL EFFECT}\footnote{Work supported in part by the
NSF:  PHY 91-23780} \\

\bigskip
\medskip
Myung-Hoon Chung \\

\medskip
{\it The Enrico Fermi Institute and the Department of Physics \\
University of Chicago,
Chicago, IL 60637}\footnote{email:  c/o lois@control.uchicago.edu}

\end{center}

\bigskip
\bigskip
\centerline{\bf Abstract}
\begin{quote}
A many-particle Hamiltonian is proposed in order to explain the
fractional quantum Hall effect (\fq)
for fractional filling factors $\nu < 1$.
The solutions of the corresponding Hartree\--Fock equations make
it possible to discuss the \fq
>from the point of view of
the single quasi\--particle energy spectrum.
It is shown how the specific couplings in the many\--particle Hamiltonian
depend on the magnetic field and the area density of electrons.
The degeneracies of the quasi\--particle states are related to the
fractional filling factors $\nu$.
It is suggested that the energy gaps obtained in the
quasi\--particle energy spectrum are comparable with the experimentally
measured quantities.
An explicit calculation for the FQH - conductance is given and its
character as a topological invariant is discussed.
\end{quote}

\vfill
\eject

\leftline{1.  INTRODUCTION}

\bigskip

It is the purpose of this paper to find a form of the
many\--particle Hamiltonian, which is relevant to the fractional quantum
Hall effect (FQHE).
In two\--dimensional electron systems subjected to huge
magnetic fields at low temperatures,
the \fq is characterized by the presence of
the plateau in the Hall resistance $\rho_{xy}$ quantized to
$(h/e^2)/(p/q)$, and the concurrent minima in the diagonal resistivity
$\rho_{xx}$.
The characteristic features of the FQHE occur at the fractional filling
factor $\nu = \frac{p}{q}$, which is by definition the ratio of the
number of electrons $\ne$ to the Landau\--level degeneracy $N$.
Apart from the quantized rational numbers $\nu = {p/q}$, other
interesting quantities measured in the \fq are the energy gaps
$\Delta_{p/q}$, which are determined by the temperature dependence of
$\rho_{xx}$.
These two quantities $\nu = {p/q}$ and $\Delta_{p/q}$
 should be explained in order to understand the \fq.

Before we proceed further to the \fq, it is instructive to consider the
classical Hall effect and the integer QHE.
It is the Drude model that reveals the overall magnetic field dependence of
$\rho_{xy}$~.
$$
\rho_{xy} = \frac{B}{nec} ~~,
\eqno(1.1)
$$
This is the essence of the classical Hall effect$^1$.
Here $n$ denotes the area density of electrons and $B$ is the magnetic
field.
The well\--known result for the degeneracy $N$ in the Landau theory is used
in order to write $\nu$  in terms of $n$ and $B$.
$$
\nu = \frac{N_e}{N} = \frac{nhc}{Be} ~~~.
\eqno(1.2)
$$
In the quantum mechanical view, we find the expression of $\rho_{xy}$ in
terms of the fundamental constants and the filling factor:
$$
\rho_{xy} = \frac{h}{e^2 \nu} ~~~.
\eqno(1.3)
$$
The plateau of $\rho_{xy}$ at integer values of $\nu$ are observed,
and are called the integer quantum Hall effect$^2$.

The nature of the integer values of $\nu$ in the IQHE is readily
understood in the view of the Landau\--level.
The completely  filled states are associated with the IQHE.
It is a generally accepted fact that the plateau of $\rho_{xy}$ are due to the
localization.
In the presence of random impurities, electrons in the system are
classified into the two kinds:
the extended states and the localized states.
Since it seems true that the completely filled states are energetically
favorable, electrons are absorbed into or supplied from the set of the
localized states in order for the system to stay in the completely filled
states.
In this sense, the magnitude of $n/B$ is stuck and the plateau appear.

Let us continue to discuss the \fq.
A few years  after the discovery of the IQHE, fractional values of
$\nu$ were  observed with a very high mobility sample$^3$.
The observed fractional filling factors$^{4,5,6}$ below
1 are expressed in terms of three quantum numbers $n, s$, and $l$
$(n \in Z_+ \equiv \{ 1,2,3 \ldots \};
s = \pm 1;
l \in Z_+ )$:
$$
\nu = \left \{
\begin{array}{l c l}
\frac{l}{2nl+s} & {\rm for} & \nu < \frac{1}{2} \\
                &          &  \\
1 - \frac{l}{2nl+s} & {\rm for} & \frac{1}{2} < \nu < 1 . \\
\end{array}
\right .
\eqno(1.4)
$$
This peculiar pattern of $\nu$ seems most likely to be the consequence of a
many\--body effect.

Laughlin$^7$  proposed a many\--body wave function to describe the \fq.
Although the Laughlin wave function incorporates the correct view of the
many\--body effect,
an explicit derivation of the wave function from a Hamiltonian is not
known.
In particular, a clear explanation of the quantum numbers in $\nu$ cannot
be found in the hierarchical constructions$^8$ based on the Laughlin wave
function.
Furthermore, the wave function approach does not uniquely determine
the energy gaps at $\nu = {p/q}$ as functions of $p$ and $q$
$^9$.

In explanations of the \fq is there another nonperturbative approach
using the topological invariant?$^{10}$
It is shown that, whenever the Fermi level lies in a gap, the Hall
conductance can be written as a topological invariant.
The meaning of the topological invariant is that the quantized value is
unchanged under small variations of interactions as long as there is
still an energy gap above the Fermi level.
It is argued in this topological approach that the many\--particle ground
state for the \fq should have the proper degeneracy.
Although this approach seems to explain the nature of
the plateaux and the
requirement of energy gaps and degeneracy, the selection rule for the
rational value of $\nu$, like for instance the odd integer in the
denominator is still missing.
Also the origin of the gaps and the degeneracy comes into question in this
approach.$^{11}$

In this paper we  introduce a many-particle Hamiltonian written in
the second quantization formalism,
presenting a quite different theoretical framework from the
wave function approach.
Starting from  the many\--particle
Hamiltonian, we  explain the pattern of $\nu = {p/q}$,
obtain the quantity of $\Delta_{p/q}$
and a reasonable value for the degeneracy of the many\--particle ground
state, with which the Hall conductance as a topological invariant is
discussed.
Our procedure  is summarized as follows.

In Sec.~2, in order to implement the idea that, when corresponding
states are completely filled, the \fq occurs
as the IQHE does, we propose a many\--particle Hamiltonian, which
modifies the Landau\--level.
We focus our attention on  the \fq with $\nu < 1$ only.
We restrict the
one\--particle states, which compose the $N$\--dimensional space, and
write the many\--particle Hamiltonian with the coupling $E(t)$.
Applying the Hartree\--Fock method, we find the
quasi\--particle energy spectrum, which is essential for the next
Section.

The spectrum is fully discussed in Sec.~3 for the case of a given
specific form of $E(t)$, where parameters $\Delta$ and $J$ are involved.
The essential results  are the
degeneracies and the energy gaps of the states, which are related to
the experimentally measurable quantities.
It is deduced how the detailed form of $J$ depends on $N$ and $N_e$.
Although the two quantum numbers $n$ and $s$ in $\nu$ are unfortunately
beyond our understanding, the origin of the quantum number $l$ is
explained quite naturally in our theory.
In addition, our theory also provides us with the energy gap
$\Delta_{p/q} = N_e \Delta/p = N \Delta /q$ at $\nu = {p/q}$.
It is also found that the degeneracy of the many\--particle ground state
at $\nu ={p/q}$ is given by $_qC_p = q!/p!(q-p)!$.
We notice that there is  symmetry breaking in the many\--particle ground
state.

By using the energy gap and the degeneracy of the many\--particle ground
state, we discuss the Hall conductance as a topological invariant in
Sec.~4.

We conclude in Sec.~5.
In order to see the consistency of the concept that  completely
filled states are associated with each $\nu$, we present a
spectrum\--like figure in the Appendix.

\bigskip
\noindent
2.  MANY-PARTICLE HAMILTONIAN AND THE \\
HARTREE-FOCK EQUATIONS

\bigskip

One of the aims in our study of the \fq is to understand the observed
rational numbers.
To do so, our first task is to derive common theoretical  properties,
which the QHE systems all share.
Presenting our conclusion first,
the essential point is that completely filled
states are associated with the QHE systems
at the observed filling factors.
In order to understand this fact, we begin by considering the
Landau\--level,  which
is used in the explanation of the IQHE.

The Hamiltonian$^{12}$ $H_{mag}$ for an electron in the magnetic field $B$ is
written as
$$
H_{mag} = \frac{1}{2} | \frac{\nabla}{i} + A |^2 ~~,
\eqno(2.1)
$$
where we set all constants to be units $m = e = \hbar = c = 1$ as usual.
Solving the corresponding Schr\"odinger equation, we impose the
twisted doubly
periodic boundary conditions
requiring that the eigenfunctions are unchanged under
translation by $L$ in the $x$ or $y$ directions up to a gauge
transformation.
The torus geometry is introduced by these conditions.
The state space is found and characterized by the Landau\--levels, where each
level has the degeneracy
$$
N = \frac{BL^2}{2 \pi} ~~.
\eqno(2.2)
$$
The energy gap between adjacent levels is  given by
$$
\Delta_{mag} = B~.
\eqno(2.3)
$$

If the first Landau\--level is
completely filled at a sample\--dependent magnetic field $B_0 = N_e 2
\pi /L^2$, then the system
at $B_0 / i$ ($i =$ integer)
with a fixed number of electrons, corresponds
to states  where electrons
are  filled up to the i$^{th}$ Landau\--level.
The Landau\--level also shows that, for the IQHE systems at the
completely filled states, there are energy gaps which are required when
an electron in the system goes to another state.
We deduce that there is no collision in the IQHE system at
the completely filled states, because there is no place to go after
collision without
paying the energy gap cost at a very low temperature.
The drift
velocity with of no collision explains the experimentally observed minima of
$\rho_{xx}$.

Although we have omitted discussing the spin effect in the above because
we are eventually interested in the region of $\nu < 1$ where the spin
effect is absent, we can conclude that the completely filled states are
associated with the IQHE.
In order to extend this idea to the \fq, we should modify the
energy spectrum.
In order to understand the fractional filling factors, we consider
a many\--particle Hamiltonian.

A master Hamiltonian $H_{tot}$ for the QHE would contain the term of
localization $H_{loc}$ as well as the term of many\--particle interaction
$H_{int}$:
$$
H_{tot} = H_0 + H_{int} + H_{loc} ~.
\eqno(2.4)
$$
The Hamiltonians $H_{int}$ and $H_{loc}$ represent
electron\--magnetic field, electron\--electron and electron\--random
impurity interactions respectively.
The observed plateaux are explained by using $H_{loc}$, which allows
flexibility of the number of the extended state electrons
carrying the current, and makes the
QHE system prefer to stay at completely filled states.
In other words, the persistance of the plateaux is attributed
respectively to the localization of excess quasi\--particles, and to the
transfer of localized states into a level in order to keep it completely
filled.
As far as the filling factors $\nu$ are concerned, we can drop the term
$H_{loc}$, making an ideal system.

In order
to discuss the idealized many\--particle Hamiltonian, $H_0 + H_{int}$, in
the second quantization formalism,
the one\--particle state space of the Hamiltonian $H_{mag}$
is considered.
We denote the states of the Landau\--level using the two quantum numbers
as $| i , j > $
with the periodicity condition
$| i,j,> = | i,j+N>$ for the i$^{th}$ Landau\--level.
One\--particle creation operators are defined as
$$
c_{ij}^\dagger  | 0 > = |i, j > ~,
\eqno(2.5)
$$
where $|0>$ is the vacuum state.
In the second quantization formalism, the term $H_0$ is written as
$$
H_0 = \sum H_{mag} = \sum_{ij} \epsilon_i c_{ij}^\dagger c_{ij} ~.
\eqno(2.6)
$$
We adopt two-particle interactions for $H_{int}$, guessing that the total
quantum numbers are preserved during the process of the interactions.
This guess is made concrete by writing $H_{int}$ in terms of the creation
operators as
$$
H_{int} = \sum_{i ~m~ r~j~ k~ t} E(r,t) c_{i+r j+t}^\dagger c_{m-r
k-t}^\dagger
 c_{mk}c_{ij} ~.
\eqno(2.7)
$$

It is enough to consider only the first Landau\--level at a very low
temperature if the degeneracy of the Landau\--level $N$ is so large that
all electrons are in the first level.
The reason is that the energy gap, proportional to the magnetic field, is
big and the Boltzmann factor for the excitation which is extremely small
at  very low temperatures.
Hence, as far as $N>N_e$, we drop the quantum number $i$, which
corresponds to the i$^{th}$ Landau\--level, in the Hamiltonian.
As a result, the following reduced many\--particle Hamiltonian $H_{red}$
is our starting point for explanation of the \fq of $\nu < 1$:
$$
H_{red} = \sum_{j=1}^N \epsilon c_j^\dagger c_j + \frac{1}{2}
 \sum_{j~ k~ t}^N E(t)
c_{j+t}^\dagger c_{k-t}^\dagger  c_k c_j~.
\eqno(2.8)
$$
Here, an unspecified value $ \epsilon$ is the energy of the first
Landau\--level.
It is remarkable that $\epsilon$ is independent of the quantum number $j$.
We will see that this independence leads to a simple calculation.

It is the Hartree-Fock method that is useful for approximating the ground
state of a system of $N_e$ interacting fermions.
The method, without detailed derivation, is used in the following discussion.
A more complete derivation can be found in many textbooks.$^{13}$

If we represent the Hamiltonian in the new one\--particle basis
$b_\alpha^\dagger$,
$$
b_\alpha^\dagger = \frac{1}{\sqrt{N}} \sum_{j=1}^N \exp
\left (i \frac{2 \pi}{N} \alpha
j\right )c_j^\dagger ~,
\eqno(2.9)
$$
the interaction term of Eq.~(2.8) is diagonalized as
$$
H_{red} = \sum_{\alpha = 1}^N \epsilon b_\alpha^\dagger b_\alpha +
 \frac{1}{2} \sum_{\alpha \beta}^N
D (\alpha - \beta ) b_\alpha^\dagger b_\beta^\dagger b_\beta b_\alpha ~,
\eqno(2.10)
$$
where the couplings $D( \alpha - \beta )$ are given by
$$
D ( \alpha - \beta ) = \sum_{t=1}^N E(t) \exp
\left (- i \frac{2 \pi}{N} t
(\alpha - \beta ) \right ) ~.
\eqno(2.11)
$$
The essential part of the method is to solve the so\--called
Hartree\--Fock equations corresponding to the eigenvalue problem.
For our case of Eq.~(2.10), the Hartree\--Fock equations are written as
$$
 \sum_{\beta=1}^N \sum_{i=1}^{N_e} \left \{ < t_i |
\beta > D (\alpha - \beta ) <\beta | t_i > < \alpha | k_j > \right .
$$
$$
\left . - < t_i | \beta > D (\alpha -\beta ) < \alpha | t_i > < \beta |
k_j > \right \}
= (\epsilon (k_j) - \epsilon ) < \alpha | k_j > ~ ,
\eqno(2.12)
$$
where the states with Greek indices are given by
$| \alpha > = b_\alpha^\dagger | 0 >$, while the states with Latin
indices $ | k_j > $ are eigenstates of the effective one\--particle
Hamiltonian.
It should be emphasized that the states $\{ | t_i > | i = 1, 2, \ldots ,
N_e \}$ must be consistently chosen from among $\{ | k_j > | j = 1, 2,
\ldots , N \}$ in such a way that the expectation value of $H_{red}$ is
minimized.
Substituting our Ansatz,
$$
< \alpha | j > = \frac{1}{\sqrt{N}} \exp
\left (-i \frac{2 \pi}{N} \alpha j \right ) ~,
\eqno(2.13)
$$
where the factor $\frac{1}{\sqrt{N}}$ is determined by the requirement of
$\sum_\alpha < l | \alpha > < \alpha | k > = \delta_{lk}$ we notice the
consistency, and obtain the solution,
that is, the quasi\--particle energy
spectrum:
$$
\epsilon (k_j) = \epsilon + N_e E(N) - \sum_{i=1}^{N_e} E (k_j - t_i )~.
\eqno(2.14)
$$
We notice from Eqs.~(2.9) and (2.13) that the effective one\--particle states
$|k_j >$ are related to the original creation operators as
$$
| k_j > = c_{k_j}^\dagger | 0 > ~.
\eqno(2.15)
$$
This fact shows that the Hartree\--Fock approximation, in our case, is
equivalent to the first order perturbation theory with the ground state
$$
| \Psi_g > = \prod_{i=1}^{N_e} c_{t_i}^\dagger | 0 > ~.
\eqno(2.16)
$$
The accidental equivalence is due to the fact that $\epsilon$
 is independent of the
quantum number $j$, that is, the high degeneracy of the Landau\--level.

In terms of the eigenvalues $\epsilon (k_j )$, the ground state energy $E_g$
is given by
$$
E_g = < \Psi_g | H_{red} | \Psi_g > = \frac{1}{2} \sum_{i=1}^{N_e} \{ \epsilon
(t_i) + \epsilon \}
\eqno(2.17)
$$
This expression provides us with the general rules for choosing the set
 $\{ t_i | i=1 , \ldots , N_e \}$.
The rules are that the individual $\epsilon (k_j)$ are minimized as much as
possible, and that the following set equation holds
$$
\{ t_i | i=1, 2, \ldots , N_e \} = \{ k_j | j = 1 , 2 , \ldots , N_e \}
\eqno(2.18)
$$
for $\epsilon (k_1) \leq \epsilon (k_r) \leq \epsilon (k_z)
\leq \ldots \leq \epsilon (k_N)~.$

An excited state of our system can be approximately described by the
state, $c_j^\dagger c_k | \Psi_g >$, where $k$ and $j$ label occupied and
unoccupied single\--particle states, respectively.
The excitation energy of the system is given by
$$
<c_j^\dagger c_k \Psi_g | H_{red} | c_j^\dagger c_k \Psi_g >
 = E_g + \epsilon (j) - \epsilon (k).
 \eqno(2.19)
$$
Since we will see that the last two terms are negligible for our specific
case, the energy gap for the excitation is $\epsilon (j) - \epsilon (k)$.

Let us make a comment about the states $c_j^\dagger |0>$
and $b_\alpha^\dagger |0>$,
and about their relation with some operators.
The doubly periodic boundary conditions discussed
in order to compactify the two\--dimensional space are
expressed by means of the so\--called magnetic translation operators $S$
and $T$, which satisfy the commutation relation,$^{14}$
$$
ST = \exp \left ( i \frac{2 \pi}{N}\right ) TS ~.
\eqno(2.20)
$$
We introduce the states $| j >$ as one of representations of the
commutation relation, setting that
$$
S | j > = \exp \left  ( i \frac{2 \pi}{N}j\right )| j > ~ ,
\eqno(2.21)
$$
$$
T | j > = | j + 1 > ~.
\eqno(2.22)
$$
If the eigenstates $|j>$ of $S$ are the one\--particle states created by
the operator $c_j^\dagger$, that is, $|j> = c_j^\dagger |0>$,
then $b_\alpha^\dagger |0>$
are eigenstates of $T$:
$$
S c_j^\dagger | 0> = \exp \left ( i \frac{\pi}{N} j \right )
 c_j^\dagger | 0 > \Rightarrow T b_\alpha^\dagger | 0 >
=\exp \left (i \frac{2 \pi}{N} \alpha \right ) b_\alpha^\dagger | 0 >~.
\eqno(2.23)
$$
For the case that the one-particles state $b_\alpha^\dagger |0>$ are the
eigenstates of $S$, the similar results are obtained.

\bigskip
\noindent
3.  QUASI-PARTICLE ENERGY SPECTRUM, J-ASSUMPTION,\\
FRACTIONAL FILLING FACTORS, ENERGY GAPS, AND \\
DEGENERACY OF THE MANY-PARTICLE GROUND STATE

\bigskip

We have studied the many-particle Hamiltonian using the Hartree-Fock
method.
It is an important result that the Hartree-Fock equations are solved
exactly and that the corresponding quasi-particle energy spectrum is obtained.
Another important argument in the previous section is that the completely
filled states are associated with the fractional filling factors.
In this section, we discuss the spectrum written in terms of the parameters
$E(t)$
of our theory, in order to explain the completely filled states at
$\nu$.
As a by\--product, the energy gap at each $\nu$ can  be calculated
and be compared with other gaps at a different $\nu$.
The degeneracy of the many\--particle ground state is also considered.

In general, the fact of $|j> = |j + N>$ shows that $E(t)$ has the
property  $E(t) = E(t+N)$, where $N$ can be treated as a number of
equally spacing points on a circle.
For convenience in later discussions, we introduce a shorthand notation
for dividing set according to residues:
$$
[r]_q \equiv \{ qm+r | m = 1,2, \ldots , \frac{N}{q} \} ~ ,
\eqno(3.1)
$$
$$
U_{r=0}^{q-1} [r]_q = \{ 1,2,3, \ldots, N \} ~.
\eqno(3.2)
$$

Before we state the  form of $E(t)$ which is  relevant to the \fq, we study
an example showing how the quasi\--particle spectrum depends on $N$ and
$N_e$.
Let us consider the case where, with a positive value $\Delta , ~E(t)$
is given by
$$
E(t) =
\left \{
\begin{array}{c l}
\Delta &{\rm for}~~t \in [0]_q \\
 0&{\rm for ~~otherwise}
\end{array}
\right .
\eqno(3.3)
$$
In dealing with the quasi-particle energy spectrum of Eq.~(2.14), our main
concern is to find the occupied states, that is, the set $\{ t_i | i   =
1,2, \ldots , \ne \}$, which minimizes the ground state energy.
The property of the algebra,
$$
a - b \in  [0]_q ~~~{\rm for} ~~a, b \in [r]_q ~,
\eqno(3.4)
$$
plays a role in understanding the fact that, for the given $E(t)$ of
Eq.~(3.3), the quasi-particles primarily prefer to occupy the states of the
same residue.
For instance, for $l \nq \leq \ne \leq (l+1) \nq$, the occupied states are
given by
$$
\{ t_i | i=1, \ldots, \ne  \} =
U_{\alpha = 1}^l [r_\alpha]_q U
\{ {t'}_i | i = 1,2, \ldots , N_e - l
{N/q} ~,~
{\rm and} ~~ {t'}_u \in [r_{l+ 1}]_q \} ~,
\eqno(3.5)
$$
We notice that there is a symmetry breaking in the many\--particle ground
state
where $r_\alpha$ is equal to one of the  $q$ different residues.
We find the energy spectrum in the form
$$
\in (k_j) = \left \{
\begin{array}{l l}
\epsilon + N_e \Delta - (N/q) \Delta & {\rm for} ~~ k_j \in
U_{\alpha = 1^l} [r_\alpha]_q \\
\epsilon + N_e \Delta - \left ( N_e - l {N/q} \right ) \Delta &
{\rm for}~~ k_j \in [r_{l+1} ] _q \\
\epsilon + N_e \Delta & {\rm for ~~ otherwise}
\end{array}
\right.
\eqno (3.6)
$$
where $E(N) = \Delta$, and the corresponding degeneracies are given by $l \nq ,
{}~ \nq$, and $N - (l+1) \nq$, respectively.
We notice that for $N_e = l {N/q}$, the excitation energy is given
by $(N/q) \Delta$.

It is easy to find that the energy gaps and the degeneracies are
independent of the choice of residues for the occupied states.
The freedom in choosing residues gives rise to a nonzero degeneracy of
the many particle ground state.
In fact, for $N_e = l{N/q}$, this degeneracy
is equal to the number of possible of combinations of
choosing $l$ out of $q$:
$$
q C_l = \frac{q!}{l! (q-l)!}
\eqno(3.7)
$$

Approaching further our problem of the \fq, we introduce a slightly
different case, where $E(t)$ is given by
$$
E(t) = \left \{
\begin{array}{c l}
\Delta & {\rm for} ~~ t \in [0]_{{N/q}} \\
0 & {\rm otherwise}
\end{array}
\right .
\eqno(3.8)
$$
The nature of the number $J$ will be discussed in detail later.
Also, for a while, we do not worry about whether or not ${N/J}$
is an integer.
An integer value of ${N/J}$ will be recovered when we discuss the
fractional filling factors

As we did in the above example for
the  $E(t)$ of Eq.~(3.8) and $lJ \leq \ne
\leq (l+1)J$,
we obtain the quasi-particle energy spectrum described in Fig.~1,
replacing $\nq$ in Eq.~(3.6) formally with $J$. .

In the hole-dominant region where $N < 2 \ne$, we manipulate the spectrum
of Eq.~(2.14):
$$
\epsilon (k_j) = \epsilon + \ne E(N) - \sum_{i=1}^N E (k_j - t_i ) +
\sum_{i=1}^{N-\ne} E(k_j - \tilde{t}_i ) ~,
\eqno(3.9)
$$
where the sets are related to each other
$$
\{ \tilde{t}_i | i = 1, 2, \ldots , N-\ne \} =
\{ 1, 2, \ldots , N \} -
\{t_i | i = 1, 2, \ldots , \ne \} ~.
\eqno(3.10)
$$
The  complementarity property implies that the original condition of Eq~(2.18)
will be satisfied if we find  a set such that
$$
\{ \tilde{t}_i | i = 1, 2, \ldots , N-\ne \} =
\{k_i | i=1, \ldots , N - N_e \}
\eqno(3.11)
$$
for $\epsilon (k_1 ) \geq \epsilon (k_2 ) \geq \ldots \geq \epsilon (k_N)$.
Considering the last term of Eq.~(3.9) for $E(t)$ of Eq.~(3.8) and $lJ \leq
N-N_e \leq (l+1)J$, we find the three energy levels
$$
\epsilon (k_j) = \left \{
\begin{array}{l l}
\epsilon + \ne \Delta - J \Delta + J \Delta & {\rm for} ~~ \in
U_{\alpha =1}^l [r_\alpha ]_{{N/J}} \\
\epsilon + \ne \Delta - J \Delta + (N - \ne - lJ ) \Delta &
{\rm for} ~~ k_j \in [r_{l+1}]_{{N/J}}  \\
 \epsilon + \ne \Delta - J \Delta + & {\rm otherwise} \\
\end{array}
\right.
\eqno(3.12)
$$
where $E(N) = \Delta , \sum_{i=1}^N E(k_j - t_i) = J \Delta$, and
the corresponding degeneracies are given by $lJ , ~ J$, and
$N-(l+1)J$, respectively (see Fig.~2).
We notice the similarity when we manipulate from
$lJ \leq N - \ne \leq (l+1) J $ to  $ N - (l+1)J \leq \ne \leq N - lJ$
and compare $\epsilon + \ne \Delta - (\ne - (N - (l+1)J)) \Delta$ in
Eq.~(3.12) with $\epsilon + \ne \Delta - (\ne - lJ) \Delta $
for $lJ \leq \ne \leq (l + ?)J$ in Eq.~(3.6).

The interesting feature of the quasi-particle energy spectrum is its
degeneracies, which are integer multiples of the number $J$.
This is a clue for explanation of the quantum number $l$ in the fractional
filling factors.
Here, describing the important part of the correct form of $E(t)$ relevant
to the \fq, we postulate how  $J$ depends on $N$ and $\ne$.
It is ``assumed'' that, in terms of $N , ~ \ne$ and the quantum number
$n \in Z_+$, the number $J$ is written as
$$
J = \left \{
\begin{array}{l l c}
|N-2n \ne | \leq \ne & {\rm for} ~~ N > 2\ne & {\rm (electron)} \\
| N - 2n (n - \ne ) | \leq N - \ne & {\rm for } ~~ \ne < N < 2 \ne & {\rm
(hole)} \\
\end{array}
\right.
\eqno(3.13)
$$
The feature of the asolute value seems to be related to the quantum
number $s = \pm 1$.
We let
$$
J = \left \{
\begin{array}{l c}
s(N-2n \ne ) & ({\rm electron}) \\
s (N-2n (N-\ne )) & ({\rm hole}) \\
\end{array}
\right .
\eqno(3.14)
$$
where, from now on, the electron and hole represent the type of dominance
as indicated in Eq.~(3.13).
We notice that the condition of inequality in Eq.~(3.13) uniquely determines
the values of $n$ and $s$ for any $N$ and $\ne$.
The above statement, in brief, is elucidated in Fig.~3, where the
corresponding $n$ and $s$ are shown.
We now call the dependence the ``$J$\--assumption''.
Actually, the $J$\--assumption is the matter that we should explain for
complete understanding of the \fq.
We guess that the number $J$ may be related with the net magnetic flux of
the system.
The circular motions of electrons under the external magnetic field
becomes the source of currents which produce the magnetic field.
In this view, each electron is considered to carry $2n$ flux quanta.
A suggestion that electrons carry $2n$ flux quanta is found in Ref.~15.
At any rate, at this stage it is not known how to prove the
$J$\--assumption,
so we shall leave it as an open
problem.

Although there are unexplained areas in the above discussion, we have set
up all the equipment for connecting completely filled states to
fractional filling factors.
We notice from the spectra of Figs.~1 and 2, that completely filled states
correspond to the following equation
$$
lJ = \left \{
\begin{array}{l c}
\ne & ({\rm electron }) \\
N - \ne & ({\rm hole}) \\
\end{array}
\right .
\eqno(3.15)
$$
Substituting Eq.~(3.14) for $J$ in Eq.~(3.15), we obtain the fractional filling
factors at the completely filled states:
$$
\frac{\ne}{N} = \nu_{nsl} = \left \{
\begin{array}{l c}
\frac{l}{2nl+s} & ({\rm electron}) \\
1 - \frac{l}{2nl+s} & ({\rm hole }) \\
\end{array}
\right .
\eqno(3.16)
$$

The spectrum shows that it becomes a two-energy level system for
completely field states, and also provides us with the energy gaps
$\Delta_{nsl}$:
$$
\Delta_{nsl} = J \Delta = \left \{
\begin{array}{l c}
\frac{\ne}{l} \Delta = \frac{N}{2nl+s} \Delta & ({\rm electron}) \\
\frac{N - \ne}{l} \Delta = \frac{\ne}{(2n-1)l+s} \Delta = \frac{N}{2nl+s}
\Delta & ({\rm hole}) \\
\end{array}
\right .
\eqno(3.17)
$$
Here, Eqs.~ (3.15) and (3.16) have been used.
In a unified expression, the energy gaps are written
$$
\Delta_{p/q} = \frac{\ne \Delta}{p} = \frac{N \Delta}{q} ~~~{\rm at} ~~~
\nu = {p/q} ~~.
\eqno(3.18)
$$
We notice $N/J = q$ from Eq.~(3.17).
As we anticipate in Eq.~(3.8), the integer value of $N/J = q$
is recovered.

Although several experiments are performed in order to measure the energy
gaps, unfortunately we cannot find any experiment which verifies the
correctness of Eq.~(3.18).
However notable data related to the energy gaps are found in Refs. ~4 and
16 as shown in Table 1.
The data in Table 1 show that $\Delta$ cannot be a constant.
We should now consider how $\Delta$ depends on $N$ and $\ne$.
For instance, we assume that $\Delta$ depends on $J$ linearly:
$$
\Delta = \delta J (\delta =~~{\rm constant}).
\eqno(3.19)
$$
Then, Eq.~(3.18) follows
$$
\Delta_{p/q} = J \Delta = \delta J^2 = \delta \left ( \frac{\ne}{p} \right
)^2 = \delta \left ( \nq \right )^2 ~.
\eqno(3.20)
$$
If $\delta$ is fixed by using one of data, then the other of data
provide us with the theoretical values, $\Delta_{2/3} = 0.432K$ and
$\Delta_{3/2} = 0.472K$, which are compared with $0.38K$ and $0.5K$
respectively.
Disorder seems to broaden the bands of the states in the spectrum of
Fig.~1.
This fact results in a reduction of the magnitude of the energy gaps.
However, in this view the value of $\Delta_{2/3} = 0.432K$ is
inappropriate,
so that more considerations of $\Delta$ or of other experiments are required.

Using the fact that nothing depends on the choices of the residues
$r_\alpha$  for the occupied states, we find the degeneracy of the
many\--particle ground state for the couplings of Eq.~(3.8).
As we discussed for the case of Eq.~(3.3), the degeneracy $d$ of the
ground state for $\nu = {p/q}$ is given by (see Eq.~(3.7))
$$
d = _qC_p = \frac{q!}{p!(q-p)!}
\eqno(3.21)
$$
The degeneracy of the many-particle ground state is one of the
conditions which is required in order to write the Hall conductance of
the \fq as a topological invariant form.

\bigskip
\noindent
4.  QUANTUM HALL CONDUCTANCE AS A TOPOLOGICAL \\
INVARIANT
\bigskip

The derivation of Eq.~(1.3) in the Introduction is problematic because it
can not be justified to  mingle the two results of classical and quantum
mechanics corresponding to Eqs.~(1.1) and (1.2) respectively.
However there is a full quantum mechanical derivation of the Hall
conductance $\sigma$ using the Kubo formula.
In fact, it is found in Ref.~10 that $\sigma$ is written in the integral
form as
$$
\sigma = \frac{e^2}{h} \frac{1}{2 \pi i} \int_0^{2 \pi}
 d \theta_x \int_0^{2 \pi}
d \theta_y \left [ < \frac{\partial \phi_0}{\partial \theta_x} |
\frac{\partial \phi_0}{\partial \theta_y} > - < \frac{\partial
\phi_0}{\partial \theta_y} |
\frac{\partial \phi_0}{\partial \theta_x} > \right ] ,
\eqno(4.1)
$$
where $| \phi_0>$ is the gauge-transformed many-particle ground state:
$$
\phi_0 = \exp \left ( - i \frac{\theta_x}{L}(x_1 + \ldots + x_{\ne}  )
-i \frac{\theta_y}{L}  ( y_i + \ldots + y_{\ne} ) \right ) \psi_g
$$
$$
\phi_o \equiv < x_1, y_1, \ldots , x_\ne , y_\ne | \phi_0 > ,
\psi_g \equiv < x_1 , y_1 , \ldots x_\ne , y_\ne | \psi_g > ~.
\eqno(4.2)
$$
For convenience, we have let the magnetic field $B$ be a unit in this
section.
The two parameters $\theta_x$ and $\theta_y$ are introduced when the
twisted doubly periodic boundary conditions are imposed on the
many\--particle ground state such as for the Landau gauge,
$$
\psi_g (x_1, \ldots , x_i + L , \ldots, y_\ne ) = \exp (i \theta_x)
\psi_g (x_1 , \ldots , x_i , \ldots , y_\ne )
\eqno(4.3)
$$
$$
\psi_g (x_1 , \ldots , y_i + L, \ldots , y_\ne ) = \exp (i \theta_y )
\exp ( - iL x_i) \psi_g (x_1 , \ldots , y_l , \ldots , y_\ne )
\eqno(4.4)
$$
It is shown in Ref.~10 that the value of $\sigma$ is always an integer
multiple of $e^2/h$ and a topological invariant as long as the
many\--particle ground state is nondegenerate, and is seperated from the
excited states by a finite energy gap.

In this section we shall calculate $\sigma$ by using $: \psi_g >$ of
Eq.~(2.16), and study the property of a topological invariant.
Although the condition of the finite energy gap is satisfied in our case
of $\nu = {p/q}$, the degeneracy of the many\--particle ground
state is given by $_qC_p$, so that we expect a noninteger value of $\sigma$.
Since the many\--particle ground state of Eq.~(2.16) in the coordinate
representation is given by the Slater determinant, it is important to
find the coordinate representation of the state created by the operator
$c_l^\dagger$.
Here, we suppose that $c_l^\dagger | 0>$ are identified as eigenstates of
the magnetic translation operator $S$.
Then we can use the solutions of $< x, y | c_l^\dagger | 0>$ found in
Ref.~12:  explicitly
$$
< x \cdot y | c_l^\dagger |0> = \exp \left ( - l/2 y^2 \right )
\frac{1}{\sqrt{L} \pi^{1 /4} } \sum_{k \epsilon Z} \exp
$$
$$
\left ( - \pi N (k + l/N + \frac{\theta_x}{2 \pi N} )^2 + i 2 \pi N
(k+ l/N + \frac{\theta_x}{2 \pi N}) (\omega /L - \frac{\theta_y}{2 \pi N}
) \right )
$$
$$
\equiv \exp ( - 1/2  y^2) f_l (\omega ; \theta_x , \theta_y )
\eqno(4.5)
$$
where $\omega = x + iy$, and $L^2 = 2 \pi N$ is assumed in order to
satisfy the condition that the total flux through the surface of the
torus is an integer in magnetic units.
The coordinate representation of the many\--particle ground state is
written as the Slater determinant with the totally antisymmetric tensor
$\epsilon_{i j \ldots k} = \pm 1$:
$$
\psi_g = \frac{1}{\sqrt{\ne !}} \exp
\left ( - 1/2 (y_1^2 + \ldots + y_\ne^2 ) \right )
$$
$$
\sum_{i_1 \ldots i_\ne}
 \epsilon_{i_1 \ldots i_\ne} f_{t_{i_1}} (x_1, y_1)
\ldots f_{t_{i_\ne}} (k_\ne , y_\ne )
\eqno(4.6)
$$
By using the properties of $f_l$:
$$
f_l (\omega + L; \theta_x , \theta_y) = \exp (i \theta_x) f_l (\omega ;
\theta_x  , \theta_y)
\eqno(4.7)
$$
$$
f_l( \omega + iL; \theta_x , \theta_y) = \exp (i \theta_y)
\exp (1/2 L^2 - i L \omega ) f_l | \omega ; \theta_x ,\theta_y)
\eqno(4.8)
$$
We notice that $\psi_g$ satisfies the twisted doubly periodic boundary
conditions of Eqs.~(4.3) and (4.4).
Futhermore, the functions $f_l$ transform for the variations of
$\theta_x$ and $\theta_y$ as follows:
$$
f_l (\omega; \theta_x  + 2 \pi , \theta_y ) = f_{l+1} (\omega ; \theta_x
, \theta_y )
\eqno(4.9)
$$
$$
f_l (\omega ; \theta_x \cdot \theta_y + 2 \pi ) = \exp (-i2 \pi l/N - i
\frac{\theta_x}{N} ) f_l (\omega ; \theta_x , \theta_y )
\eqno(4.10)
$$
As a preparation for calculation of $\sigma$, we find that
$$
\int_0^L dx \int_0^L dy \exp (-y^2) \left [ \exp \left ( - i
\frac{\theta_x}{L} x-i \frac{\theta_y}{L} y \right ) f_j \right ]^*
\left \{
\begin{array}{c}
1 \\
\frac{\partial}{\partial \theta_x} \\
\frac{\partial}{\partial \theta_y} \\
\end{array}
\right \}
$$
$$
\left [ \exp \left ( - i \frac{\theta_x}{L} x - i
\frac{\theta_y}{L} y \right ) f_l \right ] =
\left \{
\begin{array}{c}
\delta_{jl} \\
\frac{-i}{L^2} \theta_y \delta_{jl} \\
0 \\
\end{array}
\right \}
\eqno(4.11)
$$
Now let us carry out the calculation of $\sigma$
$$
\sigma = \frac{e^2}{h} \frac{1}{2 \pi i}
\int_0^{2 \pi} d \theta_x \int_0^{2 \pi} d \theta_y \int_0^L d^\ne x
\int_0^L d^\ne y
\left ( \frac{\partial \phi_0^*}{\partial \theta_x}
\frac{\partial \phi_0}{\partial \theta_y} -
\frac{\partial \phi_0^*}{\partial \theta_y}
\frac{\partial \phi_0}{\partial \theta_x} \right )
$$
$$
= \frac{e^2}{h} \frac{1}{2 \pi i}
\int_0^{2 \pi} d \theta_x \int_0^{2 \pi} d \theta_y
\left ( \frac{\partial}{\partial \theta_x} < \phi_0^*
\frac{\partial \phi_0}{\partial \theta_y} > -
\frac{\partial}{\partial \theta_y} < \phi_0^*
\frac{\partial \phi_0}{\partial \theta_y} > \right )
\eqno(4.12)
$$
where we have used the shorthand notation
$$
<K> = \int_0^L d^\ne x \int_0^L d^\ne y K ~.
\eqno(4.13)
$$
By using Green's formula:
$$
\int \int_R dxdy \left ( \frac{\partial}{\partial x} G -
\frac{\partial}{\partial y} P
\right ) =\Phi_C ( Gdy + Pdx) ~,
\eqno(4.14)
$$
we obtain
$$
\sigma = \frac{e^2}{h} \frac{1}{2 \pi i} \left [
\int_0^{2 \pi} d \theta_x < \phi_0^*
\frac{\partial \phi_0}{\phi \theta_x} >_{\theta_y=0} +
\int_0^{2 \pi} d \theta_y < \phi_0^*
\frac{\partial \phi_0}{\phi \theta_y}>_{\theta_x = 2 \pi}
\right.
$$
$$
\left.
+ \int_{2 \pi}^0 d \theta_x
< \phi_0^*
\frac{\partial \phi_0}{\phi \theta_x} >_{\theta_y=2 \pi} +
\int_{2 \pi}^0 d \theta_y < \phi_0^*
\frac{\partial \phi_0}{\phi \theta_y}>_{\theta_x = 0} \right ]
\eqno(4.15)
$$
Substituting $\phi_0$ of Eq.~(4.2) and $\psi_g$ of Eq.~(4.6) into
Eq.~(4.15), and using Eq.~(4.11), we find
$$
\sigma = \frac{e^2}{h} \frac{1}{2 \pi i} \int_{2 \pi}^0 d \theta_x
\left [ \ne \frac{-i}{L^2} \theta_y \right]_{\theta_y = 2 \pi}
= \frac{e^2}{h} \frac{\ne}{N} = \frac{e^2}{h} \nu ~.
\eqno(4.16)
$$
Here, we notice that the result of Eq.~(4.16) derived from the
quantum mechanical
point of view is identical to that of Eq.~(1.3), and that the
integral of Eq.~(4.1) is independent of the choice of the ground state.

In Ref.~(10) there is an important argument on how to find a topological
property of $\sigma$ for the \fq.
We follow the procedure given in Ref.~10 in discussing the topological
invariance of $\sigma$ below.
In the previous section, we found that the Fermi energy gap exists only at
$ \nu = p/q$ in Eq.~(3.16).
Another fact at $ \nu = p/q$ is that the many\--particle ground state
$\psi_g$ of Eq.~(2.16) is characterized by only residues of divisor $q$,
i.e. $t_i \in U_{\alpha = 1}^p [r_\alpha ]_q$.
Thus $\psi_g$, written as the Slater determinant of Eq.~(4.6) goes back
into
itself up to a phase factor for the shift of all $f_{t_i} \rightarrow
f_{t_{i +q}}$, which is obtained by the variation of $\theta_x \rightarrow
\theta_x + 2 \pi q$ and $\theta_y \rightarrow \theta_y + 2 \pi$.
Since the integral of Eq.~(4.1) is independent of the choice of the
ground state, and since $\psi_g (\theta_x + 2 \pi k , \theta_y )$ for
integer $k$ corresponds to one of the degenerated ground states,  it
is legitimate to write $\sigma$ as an average:
$$
\sigma = \frac{e^2}{hq} \frac{1}{2 \pi i}
\int_0^{2 \pi q} d \theta_x \int_0^{2 \pi} d \theta_y
\left [ <
\frac{\partial \phi_0}{\partial \theta_x} |
\frac{\partial \phi_0}{\partial \theta_y} > -
< \frac{\partial \phi_0}{\partial \theta_y} |
\frac{\partial \phi_0}{\partial \theta_x} > \right ]
\eqno(4.17)
$$
where $q$ is chosen in order to introduce the torus $0 \leq \theta_x < 2
\pi q$ and $0 \leq \theta_y < 2 \pi$, which is related to $\psi_g$.
Now we  notice that the Hall conductance is written as the integral
of Berry's curvature over the torus.
It is well known that the integral is quantized as an integer.
Furthermore, the integer must be $p$ at $\nu = p/q$ because of Eq.~(4.16).
As in the IQHE, we can attach a topological meaning to the Hall
conductance of Eq.~(4.17) in the \fq.

\bigskip
\leftline{5.  CONCLUSION}

\bigskip

We have considered the \fq as a many\--body effect.
It is assumed that the total quantum number is preserved during the
process of the interactions at the microscopic level.
The Hartree\--Fock method enables us to find the ground state of the
interacting system by solving the corresponding eigenvalue problem of the
effective one\--particle Hamiltonian.
The ground state energy is written in terms of the couplings of the
many\--particle Hamiltonian.
Extending the idea underlying the IQHE, we identify  the difference
between the states
corresponding to the fractional filling factor
$\nu$ associated with the \fq and those at $\nu$ not related to the \fq.
The filling factors correspond to the completely filled states.
In order to explain the observed fractional filling factors $\nu$, we
make the $J$\--assumption that the couplings depend on the number of states
and the number of electrons in a special way.
The quasi\--particle energy spectrum,
together with the $J$\--assumption, makes it
possible to connect the fractional filling factors to the completely
filled states.
We find the energy gaps for excitations at the fractional
filling factors.
Furthermore, we obtain a reasonable value for the degeneracy of the
many\--particle ground state, which is another condition required in
order to explain the Hall conductance of the \fq as a topological
invariant.
Using the many\--particle ground state, we carry out an explicit
calculation of
the Hall conductance $\sigma$ in the integral form, and discuss the topological
invariance.
In summary,  in order to explain the \fq, we choose
the many\--particle Hamiltonian,  the couplings,
and the nature of the one\--particle states:
Eqs.~(2.8), (3.8), (3.13) and (4.5).
The results  we derive are the fractional filling factors at the
completely filled states,  the corresponding energy gaps,
the degeneracy of the many\--particle ground state, and the Hall
conductance: Eqs.~(3.16), (3.18), (3.19) and (4.16).
There are several points which require further study.
\bigskip

(1) It is expected that there is an adequate two\--body interaction $V_2$,
 producing the $J$\--assumption in the
calculation of $< 0 | c_i c_j V_2 c_k^\dagger c_l^\dagger | 0 >$.
It is  interesting to calculate the magnitude of the parameter $\Delta$
>from the two\--body interaction.
\medskip

(2) It is important to compare the wave function $\psi_g$ in Eq.~(4.6)
with the Laughlin wave function.
Furthermore, as Arovas, Schrieffer, and Wilczek$^{17}$ did, we should
study whether or not the excitation has  anyonic properties.
\medskip

(3) Our intuition leads us to consider the $2+1$ dimensional space $(x,
y, t)$ when an action is used in order to solve problems  the \fq .
However, the torus geometry introduced by the doubly periodic conditions
seems to produce the artificial one-dimensional space where the
quasi\--particles live as shown in the Hamiltonian of Eq.~(2.8).
Thus, the $1+1$ dimensional space is involved in an action derived from
the Hamiltonian of Eq.~(2.8), that is,$ \left ( \frac{2 \pi}{N} j, t
\right )$ where $j =$ integer, $1 \leq j \leq N$.
In the continuum limit of $N \rightarrow \infty$, the corresponding space
becomes a cylindrical one:
$(l, t)$ where $0 \leq l \leq 2 \pi$.
This two\--dimensionality seems to be a clue for the connection to conformal
field theory, as far as the \fq is related to the critical phase
transition.
In two\--dimensional theory, it is well\--known that the four fermion
interaction adopted in this paper is transformed into the exponential
interaction by the bosonization.$^{18}$
Maybe this relation supports the argument that the vertex
operators$^{12}$ can be used in the explanation of the \fq.
However, it should be emphasized that the two\--dimensional space, where
the vertex operators are defined, has nothing to do with the
two\--dimensional space where real electrons are trapped in
the experiment.
\medskip

(4) As an ultimate goal in understanding of the \fq, the localization$^{19}$
should
be explained in terms of the Hamiltonian.
Progress in this area is anticipated.

\newpage

\leftline{APPENDIX}

\bigskip

We introduce a spectrum-like figure, with which all the fractional filling
factors are accompanied.
The artificial spectrum is described as the first
splitting of the Landau\--level  in
the unique way shown in Fig.~4, where the ratios of the degeneracies are
written explicitly.
For splitting the sub\--level of degeneracy $\frac{1}{1 \cdot m} $, we
use the equality:
$$
\frac{1}{1 \cdot m} =
\frac{1}{1 \cdot (m + 2)} +
\sum_{k=0}^\infty \left [
\frac{1}{ \{ m+2 + k (m+1 )\}\{m+2 + (k+1)(m+1 \} } \right .
$$
$$
\left . + \frac{1}{ \{ m+k (m+1) \} \{ m + (k+1) (m+1) \} } \right ]
\eqno(A.1)
$$

It is easy to see that the splittings follow the concept that the
completely filled states are associated with the fractional filling
factors, which are given by the sums of the degeneracies up to the
corresponding sub\--level.
For instance, if the sub\--level of degeneracy $\frac{1}{9 \cdot 13}$ is
completely filled, in other words, electrons are filled from the bottom
up to the level of $\frac{1}{9 \cdot 13}$, the following holds:
$$
\ne = \frac{1}{1 \cdot 5} N + \frac{1}{5 \cdot 9} N +
\frac{1}{9 \cdot 13} N ~,
\eqno(A.2)
$$
>from which we obtain the fractional filling factor,
$$
\nu = \frac{\ne}{N} = \frac{1}{4} ( 1 - \frac{1}{13}) = \frac{3}{13} ~.
\eqno(A.3)
$$
It is remarkable that the splittings satisfy the adiabatic theorem,
which  says that state\--dimensionality is unchanged.

\newpage
\leftline{REFERENCES}
\begin{enumerate}

\item N. W. Ashcroft and N. D. Mermin, {\it Solid State Physics}, p. 14
(Saunders College, Philadelphia, 1976).

\item K. von Klitzing, G. Dorda and M. Pepper, {\it Phys. Rev. Lett.}
{\bf 45}, 494 (1980).

\item E. C. Tsui, H. L. St\"{o}rmer, and A. G. Gossard, {\it Phys. Rev.
Lett.} {\bf 48}, 1559 (1982).

\item A. M. Chang, P. Berglund, D. C. Tsui, H. L. St\"{o}rmer, and J. C.
M. Hwang, {\it Phys. Rev. Lett.} {\bf 53}, 997 (1984).

\item R. Willet, J. P. Eisenstein, H. L. St\"{o}rmer, D. C. Tsui, A. G.
Gossard, and J. H. English, {\it Phys. Rev. Lett.} {\bf 59}, 1776 (1987).

\item H. W. Jiang, R. L. Willet, H. L. St\"{o}rmer, D. C. Tsui, L. N.
Pfeiffer, and K. W. West,{\it Phys. Rev. Lett.} {\bf 65}, 633 (1990).

\item R. B. Laughlin, {\it Phys. Rev. Lett.} {\bf 50}, 1395 (1983).

\item F. D. M. Haldane, {\it Phys. Rev. Lett.} {\bf 51}, 605 (1983).

\item B. I. Halperin, {\it Phys. Rev. Lett.} {\bf 52}, 1583 (1984).

\item Q. Niu, D. J. Thouless, and Y.-S. Wu, {\it Phys. Rev.} {\bf B31},
3372 (1985).

\item Y. S. Wu, {\it Topological Aspects of the Quantum Hall Effect},
preprint IASSNS-HEP-90/33.

\item G. Cristofano, G. Maiella, R. Musto, and F. Nicodimi, {\it Phys.
Lett.} {\bf B262}, 88 (1991).

\item For a review, see {\it Quantum Mechanics, Second Edition}, p. 535,
by E. Merzbacher (John Wiley \& Sons, Inc., New York, 1970).

\item G. 't Hooft, {\it Nucl. Phys.} {\bf B138}, 1 (1978).

\item J. K. Jain, {\it Phys. Rev. Lett.} {\bf 63} 199 (1989).

\item A. M. Chang, M. A. Paalanen, D. C. Tsui, H. L. St\"{o}rmer, and J.
C. M. Hwang, {\it Phys. Rev.} {\bf B28}, 6133 (1983).

\item D. Arovas, J. R. Schrieffer and F. Wilczek, {\it Phys. Rev. Lett.}
{\bf 53}, 722 (1984).

\item P. B. Wiegmann, {\it J. Phys.} {\bf C11}, 1583 (1978).

\item P. A. Lee and T. V. Ramakrishnan, {\it Rev. Mod. Phys.} {\bf 57},
287 (1985).

\end{enumerate}

\newpage

\leftline{FIGURE CAPTIONS}

\bigskip

FIG.~1:  The Hartree-Fock quasi-particle energy spectrum is shown for the
electron dominant region of $ N > 2\ne$.
The degeneracies and the energy gaps are given for $lJ \leq \ne \leq
(l+1)J$.
The Fermi energy is denoted as $\epsilon_F$.

\bigskip

FIG.~2:  The energy spectrum for the hole dominant region of $N < 2 \ne$
and for $lJ \leq N - \ne \leq (l+1)J$ is shown.
We notice slight differences from the spectrum of Fig.~1.

\bigskip

FIG.~3:  The $N$-dependence of the number $J$ is shown.
The corresponding quantum numbers $n$ and $s$ are chosen in order to
satisfy the condition of $J \leq \ne$ for $N > 2 \ne$ and $J \leq N -
\ne$ for $N < 2 \ne$.

\bigskip

FIG.~4:  In order to see the consistency between the adiabatic theorem and
the fact that all the known fractional filling factors are associated
with the completely filled states, the splittings are introduced.
The left one of the two numbers written beside the splitting levels is
the ratio of the degeneracy of the sub\--level to that of the unperturbed
Landau\--level, and the right one is the corresponding fractional filling
factor for the case where electrons are filled up to the sub\--level.
For $N < 2 \ne$, the fractional filling factor written here corresponds
to the case where holes are filled from the top to the sub\--level, in
other words, electrons are filled up to the next  sub\--level below.
All observed fractions below 1 can be found here.

\newpage

\begin{table}
\begin{center}

\caption{The energy gaps $\Delta_\nu$ are measured
at the magnetic field $B$ for different samples,
$X^{16}$ and $Y^4$.
The units of $kG$ and $K$ are the abbreviations of kilogauss and Kelvin
degree, respectively.}
\bigskip
\begin{tabular}{| c | c | c | c |} \hline
Sample & $B(kG)$  &  $\nu$ & $\Delta_\nu (K)$ \\ \hline \hline
   &&& \\
   & $92.5$ & $ 2/3 $ & $0.83$ \\
   &&& \\ \cline{2-4}
 X  & & & \\
    & $66.8$ & $ 2/3 $ & $0.38$ \\
	& & & \\ \hline
	&&& \\
   & $243$ &  $ 2/5 $ & $1.0$ \\
   &&& \\ \cline{2-4}
 Y &&& \\
    & $167$ &  $ 3/5 $ & $0.5 $ \\
	&&& \\ \hline
\end{tabular}
\end{center}
\end{table}

\end{document}